%% file: main_IEEE_COMM.tex
\documentclass[journal,draftcls,onecolumn,12pt,twoside]{IEEEtranTCOM}
\normalsize

\usepackage[utf8]{inputenc} 
\usepackage[T1]{fontenc}
\usepackage[colorlinks,citecolor=blue,linkcolor=blue,urlcolor=blue]{hyperref}

\usepackage{algorithmic}
\usepackage{algorithm}
\usepackage{array}
\usepackage[caption=false,font=footnotesize,labelfont=sf,textfont=sf]{subfig}
\usepackage{textcomp}
\usepackage{url}
\usepackage{verbatim}
\usepackage{graphicx}
\usepackage{cite}
\usepackage{physics}
\usepackage{amssymb, amsmath,amsfonts,bm,bbm,amsthm}
\usepackage[capitalise]{cleveref}
\let\labelindent\relax
\usepackage{enumitem}

\newtheorem{theorem}{Theorem}
\newtheorem{lemma}{Lemma}
\newtheorem{corollary}{Corollary}
\newtheorem{hypothesis}{Hypothesis}

\newtheorem{definition}{Definition}
\crefname{definition}{Definition}{Definitions}
\newtheorem{remark}{Remark}
\crefname{remark}{Remark}{Remarks}
\newtheorem{proposition}{Proposition}
\crefname{proposition}{Proposition}{Propositions}
\newtheorem{property}{Property}
\crefname{property}{Property}{Properties}

\crefname{conjecture}{Conjecture}{Conjectures}

\newcommand{\opn}[1]{\operatorname{#1}}

\newcommand{\aspas}[1]{``#1''}
\renewcommand{\vu}[1]{\ensuremath{\hat{#1}}}

\allowdisplaybreaks
\newcommand{\opa}{\vu{a}}
\newcommand{\opad}{\vu{a}^{\dagger}}

\newcommand{\opb}{\vu{b}}

\newcommand{\opnn}{\vu{n}}

\newcommand{\opq}{\vu{q}}

\newcommand{\opp}{\vu{p}}

\newcommand{\opr}{\vu{r}}

\begin{document}
	
    \title{Converging State Distributions for Discrete Modulated CVQKD Protocols}%
    
    \author{Micael Andrade Dias and Francisco Marcos de Assis%
    \thanks{Micael A. Dias and Francisco M. de Assis are with the Institute for Studies in Quantum Computing and Information Federal University of Campina Grande, Paraíba, Brazil (e-mails: micael.souza@ee.ufcg.edu.br and fmarcos@dee.ufcg.edu.br).}
        }

	\begin{center}
		\thispagestyle{empty}
		\aspas{This work has been submitted to the IEEE for possible publication. Copyright may be transferred without notice, after which this version may no longer be accessible.}
	\end{center}
	\newpage
	
	\markboth{IEEE Transactions on Communications,}%
	{Submitted paper}

    \maketitle
    
    \setcounter{page}{1}
    
    \begin{abstract}
        Consider the problem of using a finite set of coherent states to distribute secret keys over a quantum channel. It is known that computing the exact secret key rate in this scenario is intractable due to the infinite dimensionality of the Hilbert spaces and usually one computes a lower bound using a Gaussian equivalent bipartite state in the entangled based version of the protocol, which leads to underestimating the actual protocol capability of generating secret keys for the sake of security. Here, we define the QKD protocol's non-Gaussianity, a function quantifying the amount of secret key rate lost due to assuming a Gaussian model when a non-Gaussian modulation was used, and develop relevant properties for it. We show that if the set of coherent states is induced by a random variable approaching the AWGN channel capacity, then the protocol's non-Gaussianity vanishes, meaning that there is no loss of secret key rate due to the use of a Gaussian model for computing bound on the secret key rate. The numerical results show that by using a 256-QAM with Gauss-Hermite shaping, the loss of secret key rate quickly falls below $10^{-5}$ as the distance increases.
    \end{abstract}
    
	
	\IEEEpeerreviewmaketitle
    
    \input{files/SEC01}
    \input{files/SEC02}

    \input{files/SEC03}

    \input{files/SEC04}

    \input{files/SEC06}

\input{main_IEEE_COMM.bbl}
    \newpage
    \appendices
    \input{files/SEC05}

    \input{files/appendix}
    
	
\end{document}

%% file: files/SEC01.tex
\section{Introduction}\label{sec:intro}
\IEEEPARstart{Q}{uantum} key distribution protocols (QKD) are one of the first applications of quantum information theory \cite{bennett2014}. They perform the task of transmitting secret keys using quantum systems such that Alice and Bob (the legitimate parties) share identical binary sequences unknown to any other third part \cite{assche2006}. The main families of protocols are the DVQKD (Discrete Variable QKD) and the CVQKD (Continuous Variable QKD), which refer to the quantum systems used during quantum communication \cite{grosshans2002,grosshans2003,weedbrook2004,weedbrook2012,laudenbach2018}. CVQKD protocols are appealing due to its compatibility with conventional optical communication devices \cite{diamanti2015,laudenbach2018} and, in order to harvest the maximum of well developed systems, CVQKD protocols with discrete modulations have been proposed and investigated \cite{zhao2009,leverrier2009,leverrier2010,bradler2018,djordjevic2019,ghorai2019,denys2021}.

The security analysis of a QKD protocol can be done under different assumptions on the capacity of a malicious eavesdropper (Eve), providing different security levels to the protocol where Alice and Bob must use the observed data to upper bound the information Eve gained during quantum transmission \cite{assche2006,cerf2007,scarani2009,diamanti2016}. Typically, security analysis is performed under the assumptions of individual, collective, or arbitrary attacks. 
Proving security against arbitrary analysis can be quite thorny, and the common path is to show that a protocol is secure under collective attacks and then extend the security to arbitrary ones up to a negligible factor. In a general manner, under the collective attacks assumption, Alice and Bob must compute the Holevo information giving Eve's maximal knowledge on the raw key by performing an optimization over all compatible quantum channels. In the case of Gaussian modulated protocols, the maximal is attained by Gaussian quantum channels provided that Gaussian states have the property to be maximal in several functionals \cite{wolf2006,garcia-patron2006,navascues2006}.

In the case of non-Gaussian modulation, Alice and Bob do not know which quantum channel maximizes Eve's knowledge, but they can upper bound this quantity by using a Gaussian equivalent quantum state to compute the Holevo information \cite{wolf2006,assche2006,navascues2006}. Such a strategy uses the estimated covariance matrix of the bipartite shared state in the entangled-based protocol and computes the entropies for the Gaussian state with the same covariance matrix. The upper bound obtained is secure but can diminish the final secret key rate as it uses a Gaussian model to upper bound quantities relative to non-Gaussian data. An analog scenario is found in classical digital communication systems, where non-Gaussian modulation is used to transmit information over additive white Gaussian noise channels (AWGN). In this case, as well as for CVQKD, the Gaussian modulation is the one that reaches the theoretical limits for their respective tasks \cite{cover2006}. In the classical setup, optimal constellation shapes approaching the AWGN capacity have been extensively investigated \cite{wu2010,meric2015} and recent developments in DM-CVQKD protocols have shown its applications in approaching optimal secret key rates \cite{denys2021}.

In this paper, we show the conditions to the convergence of constellations of coherent states towards a reference Gaussian quantum state and explore the implications to the security analysis of discrete modulated CVQKD protocols. We define a measure of non-Gaussianity of a DM-CVQKD protocol which is interpreted as the amount of SKR lost due to using a Gaussian model to provide a lower bound to protocol with non-Gaussian modulation. We show that, under suitable conditions, the constellation non-Gaussianity measure vanishes as its size increases, making the protocol's non-Gaussianity vanish as well. We also showed the conditions in which the non-Gaussianity measure of quantum state can be extended to non-Gaussian evolutions. 

The paper is structured as follows. In Sec. \ref{sec:preliminaries} the fundamental concepts of capacity approaching constellations in classical information theory are presented for the Gaussian channel, as well as the fundamentals about Gaussian quantum states and Gaussian quantum channels. Our general structure of a discrete modulated CVQKD protocol is presented in Sec. \ref{sec:dm-cvqkd}  with prepare and measure and entangled frameworks based on QAM-like constellations. The definition of a measure of non-Gaussianity for a DM-CVQKD protocol is developed in Sec. \ref{sec:nG-cvqkd}, where we show that protocols using capacity approaching constellations (in the classical setting) result in negligible non-Gaussianity. 
Finally, in Sec. \ref{sec:conclusion}, we present our conclusions.

\subsection{Notation}
In what follows, we use the standard Dirac notation for quantum mechanics. If $A$ and $B$ are quantum systems with associated Hilbert spaces $\mathcal{H}_A$ and $\mathcal{H}_B$, respectively, $\mathcal{B}(\mathcal{H}_A)$ and $\mathcal{D}(\mathcal{H}_A)$ denote the space of bounded linear operators and the set of density operators in $\mathcal{H}_A$, respectively, with elements represented as $\vu{A}\in\mathcal{B}(\mathcal{H}_A)$ and $\vu\rho\in\mathcal{D}(\mathcal{H}_A)$. The subspace of completely positive trace preserving (CPTP) linear operators from $\mathcal{H}_A$ to $\mathcal{H}_B$ is denoted as $\mathcal{Q}(\mathcal{H}_A\rightarrow\mathcal{H}_B)$. If $\mathcal{N}$ is a quantum channel to which the quantum state $\vu\rho$ is submitted, the evolution of the state, as well as the transformations of any of its quantities, are represented by using $\overset{\mathcal{N}}{\rightarrow}$. The weak convergence of a sequence $\qty{P_n}$ of probability measures is denoted as $P_n\Rightarrow P$ and the convergence in distribution of a sequence $\qty{X_n}$ of random variables is denoted by $X_n\overset{\mathcal{D}}{\rightarrow}X$.

%% file: files/SEC02.tex
\section{Preliminaries}\label{sec:preliminaries}

In this section, we will review some basic concepts of classical and quantum information theory in order to develop some of our results. In particular, we are interested in the conditions developed in \cite{schwarte1996} to achieve the additive white Gaussian noise (AWGN) channel capacity with discrete random variables. 

    \subsection{Capacity Approaching Probability Measures for Classical Channels}
    
    In this section, we review some important results on channel capacity approaching distributions. In what follows, we use some definitions and results from \cite{wu2010,schwarte1996} to establish the basic notions that will be exploited in the QKD setting.
    
    \begin{definition}[AWGN Capacity gap \cite{wu2010}]\label{def:capacity-gap}
        Let $C(snr) = \frac{1}{2}\log(1+snr)$ be the capacity of an AWGNC and $C_m(snr) = \sup I(X; \sqrt{snr}X+N)$ be the capacity of the AWGNC restricted to a $m$-point constellation. The supremum is taken over all probability distributions of X with $|supp(P_X)|\leq m$. The minimum capacity gap of a $m$-point constellation is defined as 
        \begin{equation}\label{eq:classical-capacity-gap}
            D_m(snr) = C(snr) - C_m(snr).
        \end{equation}
    \end{definition}

    Finding optimal distributions for an $m$ point constellation is not a trivial task\footnote{In fact, in \cite{wu2010} the authors emphasize that optimal $m$ point constellations on $\mathbb{R}^2$ with arbitrary values of $m$ were still not known, even though they have showed asymptotically good constellations for the unidimensional case.} and the minimum capacity gap of \eqref{eq:classical-capacity-gap} may not be achieved for an arbitrary $m$. Let $X_n$ be a random variable under the same conditions of \Cref{def:capacity-gap} and denote by $C_{X_n}(snr) = I(X_n; \sqrt{snr}X_n+N)$ the maximum information rate for the AWGN channel with input symbols drawn from $X_n$. Then we define the capacity gap relative to the (possible) non-optimal constellation $X_n$ as
    \begin{equation}\label{eq:const-capacity-gap}
        D_{X_n}(snr) = C(snr) - C_{X_n}(snr) \geq D_m(snr).
    \end{equation}
    
    To deal with convergence of probability measures, let us switch to a measure-theoretic approach. Let $(X,d)$ and $(Y,\rho)$ be separable metric spaces with Borel $\sigma$-algebras $\mathcal{X}$ and $\mathcal{Y}$, respectively\footnote{The development with measure theory was taken from \cite{schwarte1996}. More details can also be found in \cite{gray2011}.}. A memoryless channel is defined as a probability measure (a kernel) $\nu(\cdot,\cdot): X,\mathcal{Y}\rightarrow[0,1]$ where for each $x\in X$, $\nu(x,\cdot)$ is a probability measure (conditional probability $P_{Y|X}(F|x)$) and for each $B\in\mathcal{Y}$, $\nu(\cdot, B)$ is measurable.
    
    
    \begin{definition}[Divergence]\label{def:divegence}
        Given a common measurable space $(\Omega, \mathcal{A})$ and two probability measures $P$ and $M$ on it, the divergence $D(P||M)$ is defined as
        \begin{equation}
            D(P||M) = \sup_{\mathcal{R}\in P_{\mathcal{A}}}\sum_{R\in\mathcal{R}}P(R)\log\frac{P(R)}{M(R)},
        \end{equation}
        \noindent where $P_{\mathcal{A}}$ is the collection of finite $\mathcal{A}$-measurable partitions of $\Omega$.
    \end{definition}
    
    \begin{definition}[Channel Capacity]\label{def:classical-chan-capacity}
        The channel capacity $C$ under constraint $(c,\Gamma)$ can be defined as
        \begin{equation}
            C = \sup_{\substack{Q\in\mathcal{Q} \\ \int c(x)Q(dx)\leq\Gamma}}D(PQ||Q\times M_Q)
        \end{equation}
        \noindent where $\mathcal{Q}$ is the collection of probability measures on $\mathcal{X}$.
    \end{definition}
    
    \begin{hypothesis}[Weak continuity of channels]\label{hyp:channel-cont}
        For every $x\in X$, the kernel $\nu(\cdot,\cdot)$ is weakly continuous at $x$, that is, if $x_n\rightarrow x$ ($(x_n)_{n\in\mathbb{N}}\subset X$) then $\nu(x_n,\cdot)\rightarrow\nu(x,\cdot)$. 
    \end{hypothesis}
    
    \begin{theorem}\label{th:convergen-divergence-inequality}
        Consider a sequence $(Q_n)_{n\in\mathbb{N}}$ of probability measures on $\mathcal{X}$ converging to a probability measure $Q$. If \Cref{hyp:channel-cont} holds, then $PQ_n\Rightarrow PQ$. Furthermore,
        \begin{equation}
            \liminf_{n\rightarrow\infty}D(PQ_n||Q_n\times M_{Q_n})\geq D(PQ||Q\times M_Q).
        \end{equation}
    \end{theorem}
    
    \begin{theorem}\label{th:cap-measures}
        Suppose \Cref{hyp:channel-cont} holds and $Q$ is a probability measure with $C = D(PQ||Q\times M_Q)$ and $\int c(x)Q(dx)\leq\Gamma$. Any sequence $(Q_n)_{n\in\mathbb{N}}$ of probabilities measures with $Q_n\Rightarrow Q$ and obeying the energy constraint satisfies
        \begin{equation}
            D(PQ_n||Q_n\times M_{Q_n})\rightarrow C.
        \end{equation}
    \end{theorem}
    
    Using \Cref{th:cap-measures}, it follows directly from \Cref{def:capacity-gap} that if $X_n\overset{\mathcal{D}}{\rightarrow} X$, $P_{X_n}\Rightarrow P_X$ and then $D_{X_n}(snr)\rightarrow0$.

    \subsection{Gaussian Quantum States and Gaussian Quantum Channels}\label{sec:gaussian-system}

        A continuous variable quantum system is a quantum system with an infinite-dimensional Hilbert system as state space and has as its prototype $N$ harmonic oscillators corresponding to the $N$ quantized modes of the electromagnetic field \cite{weedbrook2012}. Essentially, each mode of the system is represented by an infinite-dimensional Hilbert space $\mathcal{H}$, and the composite $N$ mode system is given by the tensor product $\mathcal{H}^{\otimes N} = \bigotimes_{i=1}^N\mathcal{H}$ and is associated with $N$ pairs of bosonic field operators $\{\opa_i,\opad_i\}_{i=1}^N$, which can be arranged in a vectorial fashion as $\bm{b} = \{\opa_1\opad_1,\ldots, \opa_N\opad_N\}^T$. 
        
        The field operators must satisfy the canonical commutation relation $[\opb_i,\opb_j] = \Omega_{ij},\,\,\,(i,j=1,2,\ldots,2N),$
        \noindent where $[\hat{A},\hat{B}] = \hat{A}\hat{B} - \hat{B}\hat{A}$ is the commutator operator and $\Omega_{ij}$ is an element of the $2N\times2N$ symplectic matrix
        \begin{equation}
            \bm\Omega = \bigoplus_{k=1}^N\omega = \mqty(\omega & &\\ & \ddots & \\ & & \omega), \,\,\,\omega = \mqty(0 & 1 \\ -1 & 0).
        \end{equation}

        Each system mode has a countable infinite and orthogonal basis $\ket{n}_{n\in\mathbb{N}}$ called the Fock basis or number states, which are associated with the number operator $\opnn=\opad\opa$ as its eigenvalues, $\opnn\ket{n} = n\ket{n}$, and with the field operators by the following relations.
        \begin{align}
            \opa\ket{n} &= \sqrt{n}\ket{n-1},\,\,\,(n\geq1),\,\,\opa\ket{0} = 0,\\
            \opad\ket{n} &= \sqrt{n+1}\ket{n+1},\,\,(n\geq0),
        \end{align}
        \noindent where $\ket{0}$ is the vacuum state.

        Bosonic operators are not Hermitian ($\opad\neq\opa$) and neither are observable. The quadrature field operators are Hermitian operators constructed by combinations of the bosonic ones,
        \begin{align}
            \opq_i = \opa_i + \opad_i,\,\,\,\opp_i = i(\opad_i - \opa_i),
        \end{align}
        \noindent which are a direct analogy of momentum and position operators in ideal quantum harmonic oscillators, also with a vectorial arrangement $\bm{\opr} = (\opq_1,\opp_1\ldots, \opq_N,\opp_N)^T$, with the commutation relation $[\opr_i,\opr_j] = 2i\Omega_{ij},\,\,\,(i,j=1,2,\cdots,2N)$.

        The state of a quantum system is represented as a unit trace positive semidefinite operator $\vu\rho:\mathcal{H}^{\otimes N}\rightarrow\mathcal{H}^{\otimes N}$ called a density operator, which carries all the information about the system. We call $\mathcal{D}(\mathcal{H}^{\otimes N})$ the space of density operators, which is convex. An arbitrary quantum state with a density matrix $\vu\rho$ in a $N$-mode system can also be represented as a quasi-probability distribution, the Wigner function. Introduce the shift (displacement) operator on the phase space, also known as the Weyl operator,
        \begin{equation}
        	\vu{D}(\bm{r}) = e^{i\bm{\vu{x}}^T\bm\Omega\bm{r}},
        \end{equation}
        \noindent where $\bm{r}\in\mathbb{R}^{2N}$, and the Fourier-Weil relation for an arbitrary bounded operator $\vu{O}$,
        \begin{equation}
            \vu{O} = \frac{1}{(2\pi)^n}\int_{\mathbb{R}^{2N}}\tr[\vu{D}(\bm{r})\vu{O}]\vu{D}(-\bm{r})\dd{\bm{r}}
        \end{equation}
        \noindent with $\dd{\bm{r}} = \dd{q_1}\dd{p_1}\cdots\dd{q_N}\dd{p_N}$ allows for an equivalent representation of a quantum state. By defining the characteristic function of $\vu\rho$ as
        \begin{equation}
            \chi_{\vu\rho}(\bm{r}) = \tr[\vu{D}(\bm{r})\vu\rho],
        \end{equation}
        \noindent we get the following representation for $\vu\rho$:
        \begin{equation}
            \vu\rho = \frac{1}{(2\pi)^n}\int_{\mathbb{R}^{2N}}\chi_{\vu\rho}(\bm{r})\vu{D}(-\bm{r})\dd{\bm{r}}.
        \end{equation}
        
        Analogously to probability distributions in classical probability theory, where the Fourier transform of a characteristic function results in a probability distribution, we can apply the Fourier transform on the quantum characteristic function of a quantum state $\vu\rho$ and get the quasiprobability distribution called the Wigner function,
        \begin{equation}
        	W_{\vu\rho}(\bm{x}) = \frac{1}{(2\pi)^{2N}}\int_{\mathbb{R}^{2N}}e^{-i\bm{\vu{x}}^T\bm\Omega\bm{r}}\chi_{\vu\rho}(\bm{r})\dd{\bm{x}}.
        \end{equation}

        As the Wigner function and the quantum characteristic function are connected via a Fourier transform and the Fourier-Weyl relation links the quantum state to its characteristic function, the density operator representation is equivalent to the representation as a quasiprobability distribution. In fact, due to a quantum version of the Bochner theorem, it is possible to define whether a function is a characteristic function of a valid quantum state \cite{hudson1974,brocker1995,parthasarathy2010,dangniam2015}.

        Like random variables in probability theory, statistical moments of quantum states are of fundamental importance. The first is the displacement vector, which is the expected value of quadrature operators,
        \begin{equation}
            \bar{\bm{r}}(\vu\rho) = \ev{\bm\opr} = \tr(\opr\vu\rho),
        \end{equation}
        \noindent and the second moment is the covariance matrix $\bm\Gamma(\vu\rho)$ with elements
        \begin{equation}
            [\bm\Gamma(\vu\rho)]_{ij} = \frac12\expval{\{\opr_i - \ev{\opr_i}, \opr_j - \ev{\opr_j}\}},
        \end{equation}
        \noindent where $\{\cdot,\cdot\}$ is the anticommutator. Whenever it is clear in the context, we remove the state specification and use only $\bar{\bm{x}}$ and $\bm\Gamma$. 
        
        Gaussian quantum states are the ones whose Wigner function assumes a Gaussian pattern, and they have the particular property of being completely described by its first and second statistical moments. We will denote by $\mathcal{G}(\mathcal{H})\subset\mathcal{D}(\mathcal{H})$ the set of Gaussian states and by $\opn{Conv}(\mathcal{G})$ its convex hull\footnote{The set of Gaussian states of a quantum system is not convex and taking its convex hull may be a mathematical convenience in some applications, such as in quantum resource theories.}. Note that this comes from the fact that $\mathcal{G}(\mathcal{H}_A)$ is a proper subset of $\opn{Conv}(\mathcal{G})$, which means that, in general, convex combinations of Gaussian states do not result in Gaussian states. 
        
        Denote by $\mathcal{L}(\mathcal{H}_A,\mathcal{H}_B)$ the space of all linear transformations from system $A$ to system $B$. The evolution of a quantum system can be modeled using a completely positive trace preserving (CPTP) linear map $\mathcal{N}:\mathcal{D}(\mathcal{H}_A)\rightarrow\mathcal{D}(\mathcal{H}_B)$, which is called a quantum channel. We denote by $\mathcal{Q}(\mathcal{H}_A\rightarrow\mathcal{H}_B)\subset\mathcal{L}(\mathcal{H}_A,\mathcal{H}_B)$ the subspace of all quantum channels. We say that a CPTP reversible quantum operation is Gaussian when it transforms Gaussian states into Gaussian states. Such unitaries are generated by a Hamiltonian $\vu{H}$ that is at most a second-order polynomial on the canonical operators $\vu{U} = \exp\{-i\vu{H}/2\}$. Once again, we denote the set of Gaussian operations by $\mathcal{G}(A\rightarrow B)\subset\mathcal{Q}(\mathcal{H}_A\rightarrow\mathcal{H}_B)$.
        
        The action of any Gaussian evolution can be defined in terms of transformation of quadrature field operators as the affine map         
        \begin{equation}
        \bm{\vu{r}} \rightarrow \bm{S}\bm{\vu{r}} + \bm{d},
        \end{equation}        
        \noindent where $\bm{S}$ is a $2N\times 2N$ real symplectic matrix and $\bm{d}\in\mathbb{R}^{2N}$ with the property $\bm{S} \bm\Omega\bm{S}^T = \bm\Omega$. With respect to the statistical moments $\bar{\bm{r}}$ and $\bm\Gamma$ of an arbitrary quantum state $\vu\rho$, the action of the Gaussian evolution performs the following transformation
        \begin{align}
            \bar{\bm{r}} \rightarrow \bm{S}\bar{\bm{r}} + \bm{d}, & & \bm\Gamma\rightarrow\bm{S}\bm\Gamma\bm{S}^T,
        \end{align}
        \noindent which is valid even if $\vu\rho\notin\mathcal{G}(\mathcal{H})$.

        An important Gaussian quantum channel to practical implementation of quantum communication is the thermal-loss channel, which is the model to transmission of light through optical fibers and free space. This channel is cahracterized by the transmittance $\tau$ and thermal noise $\varepsilon = 2\bar{n}+1$ ($\bar{n}$ is the average number of thermal photons) and transforms the statistical moments as
        \begin{align}
            \bar{\bm{r}} \rightarrow \sqrt{\tau}\bar{\bm{r}}, & & \Gamma\rightarrow \tau\bm\Gamma + (1-\tau)\varepsilon\bm{I},
        \end{align}
        \noindent where we can use the change of variable $\xi = 2\bar{n}(1-\tau)$, where $\xi$ is called the channel excess noise, the average number of photons added to the system mode by coupling with the environment in a thermal state.
    
        \subsection{The non-Gaussianity of a Quantum State}

        The \aspas{Gaussian sector} is of particular importance once states and Gaussian operations are relatively easy to implement in the laboratory using lasers, amplifiers, and common optical telecommunications devices.
        In the study of Quantum Resource Theory (QRT), they may be considered to have no coast to implement, as non-Gaussian (nG) states and operations can be more challenging to prepare and realize \cite{lami2018,chitambar2019}. Several quantum information processing tasks treat nG states and operations as resources, and a whole field has been built around this idea \cite{genoni2008,genoni2010,albarelli2018,baek2018,takagi2018,zhuang2018,lee2019}.

        Within this idea of non-Gaussianity as a resource and foccusing in quantum states, it makes sense to ask how non-Gaussian a nG state is. As there is no axiomatic definition of \aspas{Gaussianity}, several functions have been proposed as measures of non-Gaussianity for quantum states. For the following definition, we need the concept of Gaussian equivalent of a quantum state: if $\vu\rho$ is an arbitrary quantum state, $\vu\rho^G$ is the Gaussian state with the same first and second statistical moments. Note that since $\vu\rho^G$ is Gaussian, it is completely described by its statistical moments and that $\vu\rho$ and $\vu\rho^G$ are not the same states, unless $\vu\rho$ is already Gaussian.
        
        \begin{definition}[QRE non-gaussianity \cite{genoni2008}]\label{def:qre-ng}
        Let $\vu\sigma$ be an arbitrary quantum state and $\vu{\sigma}^G$ its Gaussian equivalent. A measure of non-Gaussianity of $\vu\sigma$ based on the quantum relative entropy is defined as 
        \begin{equation}\label{eq:qre-ng}
            \delta_{vN}(\vu\sigma) = S(\vu\sigma||\vu\sigma^G).
        \end{equation}
        \end{definition}

        Here, $S(\vu\rho||\vu\sigma) = tr[\vu\rho(\log\vu\rho - \log\vu\sigma)]$ is the quantum relative entropy of the states $\vu\rho$ and $\vu\sigma$. Once in \eqref{eq:qre-ng} $\vu\sigma$ and $\vu\sigma^G$ have the same covariance matrix and $\log\vu\rho^G$ is an order two polynomial operator in the canonical operators, one has that $\tr[(\vu\rho-\vu\rho^G)\log\vu\rho^G] = 0$ and that $\delta_{vN}(\vu\sigma) = S(\vu{\sigma}^G) - S(\vu{\sigma})$, where $S(\vu\sigma) = -\tr(\vu\sigma\log\vu\sigma)$ is the von Neumann entropy of the quantum state $\vu\sigma$. The entropy of a $N$ mode Gaussian quantum state can be computed from the bosonic entropic function, $S(\vu\sigma^G) = \sum_{i=1}^Ng(\lambda_i)$ with
        \begin{equation}
            g(\lambda) = \qty(\frac{\lambda+1}{2})\log(\frac{\lambda+1}{2}) - \qty(\frac{\lambda-1}{2})\log(\frac{\lambda-1}{2}),
        \end{equation}
        \noindent where $\{\lambda_i\}$ is the set of symplectic eigenvalues of $\bm\Gamma(\vu\sigma^G)$. The entropy of a mixture of pure states can be obtained by computing the entropy of the corresponding Gramm matrix, as explained in Appendix \ref{ap:ent-gram-matrix}.
        
        Two relevant properties of the von Neumann nG measure are its non-negativity and contractivity under Gaussian quantum channels:

        \begin{property}\cite[Lemma 01]{genoni2008}\label{prop:ng-nonegativity}
            $\delta_{vN}(\vu\sigma)$ is a nonnegative quantity with $0\geq\delta_{vN}(\vu\sigma)\geq\infty$ and $(\vu\sigma) = 0$ if and only if $\vu\sigma$ is a Gaussian state.
        \end{property}
        \begin{property}\cite[Lemma 07]{genoni2008}\label{prop:ng-monotonicity}
            $\delta_{vN}(\vu\sigma)$ monotonically decreases under Gaussian quantum channels, that is, $\delta_{vN}(\vu\sigma) \geq \delta_{vN}(\mathcal{N}(\vu\sigma))$ for every $\mathcal{N}\in\mathcal{G}$.
        \end{property}

        The proof of the above properties can be found in \cite{genoni2008,genoni2010}.

%% file: files/SEC03.tex
\section{Secret Key Rates for Discrete Modulated CVQKD Protocols}\label{sec:dm-cvqkd}

In this section, we discuss the discrete modulated (DM) CVQKD protocol and the framework of our analysis. Consider a complex-valued random variable $X_n$, $n\geq1$ with alphabet $\mathcal{X}_n$ and $Pr[X_n = x] = p_{X_n}(x)$. For simplicity and compatibility with the usual digital communication systems, we assume that $|\mathcal{X}_n| = (n+1)^2=N$ and whenever it is clear from the context, we drop the subscript and use only $p(x_i)$. We also use the notation $[N] = \qty{1,2,\cdots,N}$ and assume that $X_n$ is symmetric around the origin\footnote{This assumptions comes without loss of generality as the constellations approaching the AWGN channel capacity are usually symmetric \cite{wu2010}.}, that is, $p(x_i) = p(-x_i)$. The $m^2$-state DM-CVQKD prepare and measure protocol induced by $X_n$ works as follows.

\begin{enumerate}[label=(\roman*)]
    \item\textit{State Preparation -} At each round, Alice draws $x$ from $X_n$ and prepares a coherent state $\ket{x}$. The modulation scheme is represented by the ensemble $\mathcal{A} = \qty{\ket{x_k}, p(x_k)}_{k=1}^N$, $x_k\in\mathcal{X}_n$, which is the mixture $\vu\rho_{X_n} = \sum_{x\in\mathcal{X}}p(x)\op{x}$ and the register $\bm{X}'$ stores Alice's random experiment results.
    
    \item\textit{Quantum Transmission and Measurement -} The prepared state is sent through a one-mode quantum channel $\mathcal{N}_{A\rightarrow B}$ such that Bob observes the mixture $\vu\rho_B = \mathcal{N}_{A\rightarrow B}(\vu\rho_{X_n})$ and performs heterodyne detection, denoted by the operator $\mathcal{M}_{B\rightarrow Y}$. The measurement results are stored in $\bm{Y}' = \mathcal{M}_{B\rightarrow Y}(\vu\rho_B)$.
    
    \item \textit{Sifting - }After the conclusion of $N$ rounds, Alice and Bob agree on a small random subset of $I_{test}\subset[N]$ to form the test set to be used in parameter estimation. The values of $\bm{X}'$ and $\bm{Y}'$ indexed by $I_{test}$ are publicly announced and then discarded. The remaining raw key values are indexed by $I_{key}=[N]\setminus I_{test}$ and represented by $\bm{X}$ and $\bm{Y}$ on the Alice and Bob sides, respectively.
    
    \item\textit{Parameter estimation - } Once the test set is defined, Alice and Bob use it to estimate the quantum channel parameters, mainly by estimating the first and second moments of the data. Based on the estimated values, they evaluate whether it is possible to distill a secret key and if it is not, they abort the protocol and go back to step (i).
    
    \item\textit{Data Estimation - } Before information reconciliation, Bob performs an estimation procedure in order to retrieve a sequence $\hat{\bm{X}} = \theta(\bm{Y})$, where $\theta$ is some estimation function. If $X_n$ has uniform distribution, that is, $p(x_i) = \frac{1}{N}$, $\theta$ is a minimum distance estimator
    \begin{equation}
        \hat{\bm{X}}[j] = \arg\min_{x\in\mathcal{X}_n}||\bm{Y}[j] - x||^2,
    \end{equation}
    \noindent where $\hat{\bm{X}}[j]$ (respec. $\bm{Y}[j]$) denotes the $j$-th element of $\hat{\bm{X}}$ (respec. $\bm{Y}$). In the case of nonuniform distribution, which will be the relevant case, $\theta$ is the \textit{maximum a posteriori} (MAP) estimator,
    \begin{equation}
        \hat{\bm{X}}[j] = \arg\max_{x\in\mathcal{X}_n}p_{X_n|Y}(x|\bm{Y}[j]),
    \end{equation}
    
    \item\textit{Reconciliation and Privacy Amplification - } Alice and Bob choose a suitable error correction protocol and privacy amplification, which must agree with the quantities evaluated during parameter estimation. They then apply error correction and privacy amplification to generate the secret key.
\end{enumerate}

\begin{figure*}[!t]
	\centering
	\subfloat[]{\includegraphics[width=2.5in]{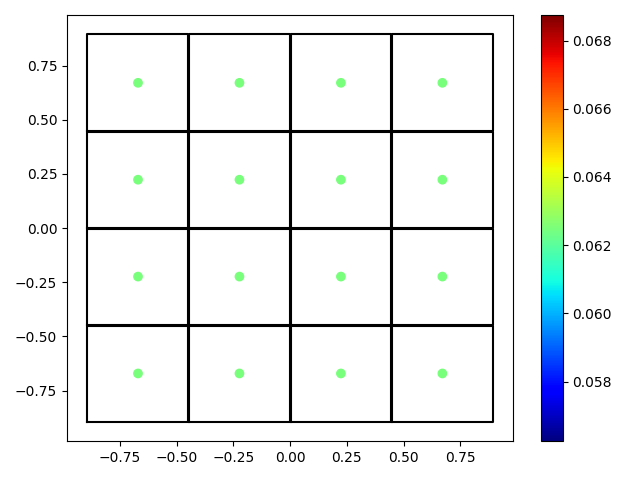}}%
	\subfloat[]{\includegraphics[width=2.5in]{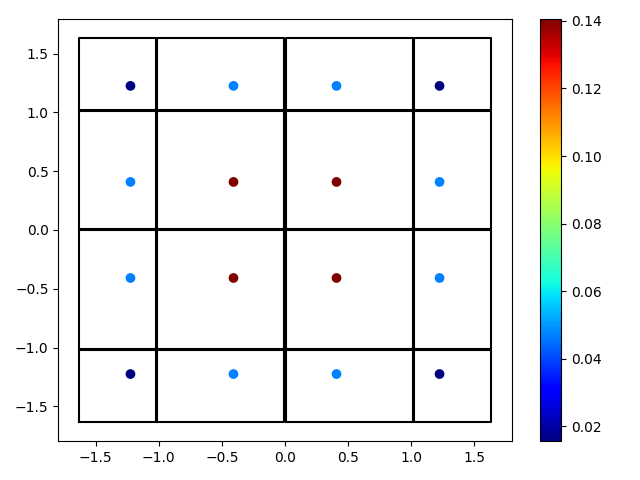}}%
	\caption{\label{fig:decision_regions}Decision Regions for equiprobable (a) and non-equiprobable (b) 16-QAM constellations.}
\end{figure*}

It is worth highlighting that, analogously to \cite{lin2019} but more generally, the constellation represented by the alphabet of $X_n$ is a set of points on the complex plane and the estimation function $\theta$ creates different decision regions for different distributions of $X_n$. In Fig.\ref{fig:decision_regions} the decision regions for equiprobable and non-equiprobable 16-QAM constellations are exemplified. Such regions depend not only on the distribution of $X_n$ but also on the channel model. If it is an additive white Gaussian noise model (AWGN), for example, the decision regions of the MAP estimator also depend on the noise variance.

Here, the relevant parameters chosen by Alice for the quantum state transmission stage are the constellation size $m^2$ and the constellation energy 
\begin{equation}
    \ev\opnn = \tr(\opad\opa\vu\rho_{X_n}) = \sum_{x\in\mathcal{X}_n}p(x)|x|^2 = \opn{Var}(X_n).
\end{equation}
\noindent which we call the modulation variance $V_m$. It is clear that $\bar{\bm{r}}(\vu\rho_{X_n}) = \bm{0}$ ($X_n$ has zero mean) and the covariance matrix $\bm\Gamma(\vu\rho_{X_n})$ is defined by
\begin{equation}
    \bm\Gamma(\vu\rho_{X_n}) = \mqty( \ev{\opq^2} & \frac12\expval{\{\opq,\opp\}} \\ \frac12\expval{\{\opq,\opp\}} & \ev{\opp^2} ),
\end{equation}
\noindent where $\frac12\expval{\{\opq,\opp\}} = 0$ and $\ev{\opq^2} = \ev{\opp^2} = 1 + 2V_m$.


    \subsection{The Equivalent Entangled Based Protocol}\label{sec:eb-protocol}
    
    Security analysis is in general more accessible when the protocol is defined upon bipartite entangled states rather than one-mode quantum systems, and such entangled-based (EB) protocol must keep an equivalence with the PM version. This equivalence is taken in the sense that any security analysis performed on the EB version is true for the PM version. 
    
    In the EB protocol equivalent to the PM protocol, in which Alice prepares coherent states from the ensemble $\mathcal{A}$ induced by the random variable $X_n$, Alice has access to a bipartite state $\ket{\Phi}_{AA'}$ that is a purification of $\vu\rho_{X_n}$, ensuring that $\vu\rho_{X_n} = \tr_{A'}(\op{\Phi}_{AA'})$. She measures her mode $A$, which has the effect of collapsing the second mode $A'$ into one of the states in\footnote{For more details about the purification of mixtures of coherent states, see \cite{leverrier2009,ghorai2019,denys2021}.} $\mathcal{A}$. This second mode is sent to Bob through the quantum channel $\mathcal{N}_{A'\rightarrow B}$, and Bob receives the output mode $B$. The density operator representing the shared bipartite state is $\vu\rho_{AB} = (\mathbbm{1}_A\otimes\mathcal{N}_{A'\rightarrow B})(\op{\Phi}_{AA'})$. 
    
    Assuming that the eavesdropper performs a collective attack over the transmitted quantum states, her machinery is capable of interacting individually with each transmitted state and having access to a quantum memory, allowing a delayed collective measurement. This collective attack strategy grants the eavesdropper the possibility of achieving the Holevo bound on the accessible information. By so, the Devetak-Winter bound on the distillable secret key rate reads \cite{devetak2005,devetak2005a}
    \begin{equation}\label{eq:dw-bound}
        K = I(X_n;Y) - \sup_{\mathcal{N}_{A'\rightarrow B}}\chi(Y;E),
    \end{equation}
    \noindent where $I(X,Y)$ is the mutual information between Alice and Bob's classical data, $\chi(Y;E)$ is the Holevo bound on Eve's accessible information of Bob's data and the supremum is over all quantum channels compatible with the observed statistics during parameter estimation. The register $E$ concerns the eavesdropper system, which can be fitted in the original model by defining the Stinespring dilation $\mathcal{U}_{A'\rightarrow BE}$ of $\mathcal{N}_{A'\rightarrow B}$. such that the state
    \begin{equation}
        \vu\rho_{ABE} = (\mathbbm{1}_A\otimes\mathcal{U}_{A'\rightarrow BE})(\op{\Phi}_{AA'})
    \end{equation}
    \noindent is a purification of $\vu\rho_{AB}$.
    
    \subsection{The Gaussian Extremality Property}\label{sec:get}
    
    The expression in \eqref{eq:dw-bound} involves maximizing the eavesdropper accessible information by searching the set of quantum channels that agree with the estimated parameters in the PM protocol. It is important to stress that the goal of a QKD protocol's security analysis is to make assumptions about the eavesdropper capabilities in order to estimate how much secret bits can be reliably distilled for the final key. Although what Alice and Bob really have are correlated sequences, and, from a practical point of view, every interaction the quantum states are submitted to is flagged as the action of a quantum channel coupling the states sent to the environment, and such a loss of information to the environment is indeed assumed to be information acquired by the eavesdropper.
    
    The extremality property of Gaussian states ensures that they are maximal for several quantities given fixed first and second moments, including von Neumann entropy and the Holevo quantity. This means that, for the protocol defined in\footnote{In fact, it is valid for an arbitrary QKD protocol using a bipartite state $\vu\rho$ with secret key rate denoted by $K(\vu\rho)$ under collective attacks assumptions.} Sec. \ref{sec:eb-protocol}, the Gaussian state $\vu\rho_{AB}^G$ \textit{equivalent} to $\vu\rho_{AB}$ provides an upper bound on $\chi(Y;E)$ and by so, a lower bound on the secret key rate, meaning that $K(\vu\rho_{AB})\geq K(\vu\rho_{AB}^G)$, which is the optimality of Gaussian attacks \cite{garcia-patron2006,navascues2006,wolf2006}.

%% file: files/SEC04.tex
\section{The non-Gassianity of DM-CVQKD Protocols}\label{sec:nG-cvqkd}

In the last Section, the DM-CVQKD protocol was described in both PM and EB implementations, and the secret key rate formula was derived for a bipartite state $\vu\rho_{AB}$, which can be lower bounded by using the equivalent Gaussian state $\vu\rho_{AB}^G$. Such a lower bound, although secure, can result in underestimating the protocol secret key rate by overestimating the eavesdropper information. This amount of SKR lost in the security analysis can be interpreted as a penalty to the protocol due to the discrete modulation used and we address this quantity as a non-Gaussianity measure of the protocol.
In this section, we develop the definitions and fundamental results about this measure of non-Gaussianity and its consequence to the security analysis.

\begin{definition}[DM-CVQKD non-gaussianity]\label{def:qkd-protocol-ng}
    Let $\vu\rho_A = \sum_ip_i\vu\rho_i$ be a non-Gaussian mixture of $N = (n+1)^2$ coherent states induced by a random variable $X_n$ as in Sec. \ref{sec:dm-cvqkd}, $\vu\rho_{AA'} = \op{\Psi}_{AA'}$ its purification, $\vu\rho_{AB} = (\mathbbm{1}_A\otimes\mathcal{N}_{A'\rightarrow B})(\op{\Phi}_{AA'})$ the shared state after sending it through the quantum channel, and $\vu\rho^G_{AB}$ the Gaussian equivalent of $\vu\rho_{AB}$. Denote by $K(\vu\rho_{AB})$ and $K(\vu\rho^G_{AB})$ the SKR for $\vu\rho_{AB}$ and $\vu\rho_{AB}^G$, respectively, under the same assumptions (channel parameters and modulation energy). The protocol's non-Gaussianity parameter defined by
    \begin{equation}
        \varepsilon_G(\vu\rho_{AB}) = K(\vu\rho_{AB}) - K(\vu\rho^G_{AB}),
    \end{equation}
    \noindent represents the loss in the secret key rate by assuming a Gaussian equivalent model of the protocol. 
\end{definition}

\begin{lemma}\label{lemma:qkd-protocol-ng-nonegativity}
    The non-Gaussianity parameter $\varepsilon_G(\vu\rho_{AB})$ is non-Negative for any DM-CVQKD protocol under collective Gaussian attacks.
\end{lemma}

\begin{proof}
    Due to the Gaussian extremality theorem, $K(\vu\rho_{AB}) \geq K(\vu\rho^G_{AB})$.
\end{proof}

\begin{lemma}\label{lemma:qkd-protocol-ng}
    For a CV-QKD protocol with non-Gaussian modulation of coherent states represented by the density operator $\vu\rho_A$, one has that
    \begin{equation}
        \varepsilon_G(\vu\rho_{AB}) = \inf_{\mathcal{N}_{A'\rightarrow B}}\qty{\delta_{vN}(\mathcal{N}(\vu\rho_A))} - D_{X_n}(\opn{snr}),
    \end{equation}
    where $\opn{snr} = \tau V_m/(1+\xi)$.
\end{lemma}

\begin{proof}
    By developing the expression for the non-Gaussianity of the protocol, we obtain the following result.
	\begin{align}
		\varepsilon_G(\vu\rho_{AB}) &= I(A;B) - \sup_{\mathcal{N}_{A'\rightarrow B}}\qty{\chi(B;E)} - [I(A^G;B) - \chi(B^G;E)]\\
		&= \chi(B^G;E) - \sup_{\mathcal{N}_{A'\rightarrow B}}\qty{\chi(B;E)} - [I(A^G;B) - I(A;B)].
	\end{align}
    \noindent where we used the superscript \aspas{G} to indicate the system's Gaussian counterpart when using the equivalent Gaussian state. The supremum does not appear in the Holevo quantity regarding the equivalent Gaussian quantum system because the Gaussian interactions are already optimal in this case and $\chi(B^G;E)$ can be computed from the covariance matrix reconstructed with the estimated parameters. As the quantum channels considered in the optimization must be compatible with the observed data, we can affirm that $\chi(B^G;E)$ depends only on the choice of $\vu\rho_{AB}$ and we can drop the supremum for this term. The term in brackets is exactly equal to the capacity gap $D_{X_n}(snr)$ of a $N$-point constellation to the AWGN channel with SNR given by $\tau\tilde{V}_m/(1+\xi)$. The difference of Holevo quantities can be developed as
    \begin{align}\nonumber
		\chi(B^G;E) - \sup_{\mathcal{N}_{A'\rightarrow B}}\chi(B;E) &= \inf_{\mathcal{N}_{A'\rightarrow B}}\qty{\chi(B^G;E) - \chi(B;E)}\\
				&= \inf_{\mathcal{N}_{A'\rightarrow B}}\qty{S(\rho^G_{AB}) - S(\rho^G_{AB}|B) - \qty[S(\rho_{AB}) - S(\rho_{AB}|B)]}\\
				&=  \inf_{\mathcal{N}_{A'\rightarrow B}} \delta_{vN}(\rho_{AB}) - \qty[\delta_{vN}(\rho_{AB}) - \delta_{vN}(\rho_B)]\\
				&= \inf_{\mathcal{N}_{A'\rightarrow B}}\delta_{vN}(\rho_B).
	\end{align}
    \noindent where in the second equality we used the identity $S(\rho^G_{AB}|B) - S(\rho_{AB}|B) = \delta_{vN}(\rho_{AB}) - \delta_{vN}(\rho_B)$. Since $\vu\rho_B = \mathcal{N}(\vu\rho_A)$, it follows that
    \begin{equation}
        \varepsilon_G(\vu\rho_{AB}) = \inf_{\mathcal{N}_{A'\rightarrow B}}\qty{\delta_{vN}(\mathcal{N}(\vu\rho_A))} - D_{X_n}(snr).
    \end{equation}
\end{proof}

What is curious about this equivalent description of the DM-CVQKD protocol's nG measure is that the difference of Holevo informations, which computes how far away the lower bound given by using the extremal properties of Gaussian state is from the actual upper bound (the supremum), equals the minimum measure of non-Gaussianity of the state observed by Bob. In other words, in a DM-CVQKD protocol, the quantum channel giving the eavesdropper the maximum information about Bob's system is the one that leaves it more like a Gaussian state.

In the point of view of computing bounds to the SKR, it is much more convenient to assume the Gaussian model to the observed data and use the extremal property of Gaussian states than to compute the supremum on the Holevo information. The penalty of the easier way is $\varepsilon_G$. In the following, we show how $\varepsilon_G$ becomes negligible with the increasse of the cardinality of suitable constellations used by Alice.

    \subsection{Convergence of Ensembles of Coherent States}

    Once the nG measure of a DM-CVQKD protocol $\varepsilon_G$ is upper bounded by the nG measure of the state observed by Bob at the channel output, it is reasonable to ask if there is any kind of constellation that Alice can use such that the state on Bob's side is \aspas{Gaussian enough}. Since the von Neumann entropy is a quantity that is maximized by Gaussian quantum state (with an energy constraint), we use as motivation the conditions of achievability of the AWGN channel capacity by probability measures converging to the normal one and ask: what if such constellations were used by Alice to transmit coherent states? Are those constellations minimizing $\varepsilon_G$? 
    
    In the following, we show some important convergence results for ensembles of coherent states induced by converging sequences of random variables. We begin by the definition of convergence of bounded operators, which allows for a result on the von Neumann entropy bound of a quantum state.

    \begin{definition}[Weak Convergence of Bounded Operators \cite{holevo2019}]\label{def:wak-conv-operators}
        A sequence of bounded operators $\vu{A}_n$ in $\mathcal{B}(\mathcal{H})$ weakly converges to an operator $\vu{A}$ if $\lim_{n\rightarrow\infty}\matrixel{\psi}{\vu{A}_n}{\phi} \rightarrow \matrixel{\psi}{\vu{A}}{\phi}$ for every $\psi,\phi\in\mathcal{H}$.
    \end{definition}
    
    \begin{theorem}[von Neumann Entropy Lower Semicontinuity \cite{holevo2019}]\label{th:entropy-semicontinuity}
        The quantum entropy is lower semicontinuous on the space of all density operators in $\mathcal{H}$. Let $\qty{\vu\sigma_n}$ be a sequence of density operators in $\mathcal{D}(\mathcal{H})$ converging to a density operator $\vu\sigma$. Then,
        \begin{equation}
            S(\vu\sigma) \leq \liminf_{n\rightarrow\infty} S(\vu\sigma_n).
        \end{equation}
    \end{theorem}
    
    For the following result, we use the convergence in distribution of a random variable towards a normal distribution to prove the convergence of (suitable) density operators towards a Gaussian state. Let $\qty{X_n}_{n\in\mathbb{N}}$ be a sequence of discrete random variables with alphabets $\mathcal{X}_n$ over the complex field and distributions $P_{X_n}$, $|\mathcal{X}_n|=(n+1)^2= N$, $\mathbb{E}[X_n]=0$ and $\mathbb{E}[X_n^2] = V_m$. We denote by $\vu\rho_{X_n}$ the ensemble of coherent states induced by the random variable $X_n$, that is, $\vu\rho_{X_n} = \sum_{i=1}^Np_{X_n}(x_i)\op{x_i}$. Analogously, if $X$ is a complex-valued continuous random variable with distribution $p_X(x)$, $\vu\rho_X = \int_\mathbb{C}p_X(x)\op{x}\dd[2]{x}$.
    
    \begin{theorem}\label{th:state-convergence}
        If $X_G\sim\mathbb{C}\mathcal{N}(0,V_m)$ and $\qty{X_n}_{n\in\mathbb{N}}$ is a sequence of random variables such that $X_n\overset{\mathcal{D}}{\rightarrow}X_G$ then $\vu\rho_{X_n}\rightarrow\vu\rho_{X_G}=\vu\rho^{th}(V_m)$.
    \end{theorem}
    
    \begin{proof}
        First, we note that $\vu\rho^{th}(V_{mod}) = \vu\rho_{X_G}$ comes from the 
        Glauber-Surdashan P-representation of a thermal state: 
        \begin{equation}
            \vu\rho^{th}(V_{mod}) = \frac{1}{\pi V_m}\int_\mathbb{C}\exp{-|x|^2/V_m}\op{x}\dd[2]{x}.
        \end{equation} 
        
        \noindent To prove the convergence, let us take arbitrary $\ket{\phi},\ket{\psi}\in\mathcal{H}$. From the definition of $\vu\rho_{X_n}$, we have that
        \begin{equation}
			\matrixel{\psi}{\vu\rho_{X_n}}{\phi} = \bra{\psi}\qty(\sum_{i=1}^np_{X_n}(x_i)\op{x_i})\ket{\phi}= \sum_{i=1}^n p_{X_n}(x_i)\braket{\psi}{x_i}\cdot\braket{x_i}{\phi},
		\end{equation}
        \noindent for which, assuming without loss of generality that $\phi = \sum_{j=0}^{\infty}\phi_j\ket{j}$ and $\psi = \sum_{k=0}^{\infty}\psi_k\ket{k}$ are the respective expansion of the states in the Fock basis, we obtain
        \begin{align}
            \braket{\psi}{x_i} = e^{-|x_i|^2/2}\sum_{j=0}^\infty\psi_j^*\frac{x_i^j}{\sqrt{j!}}, & & \braket{x_i}{\phi} = e^{-|x_i|^2/2}\sum_{k=0}^\infty\phi_k\frac{(x_i^*)^k}{\sqrt{k!}}.
        \end{align}
        
        \noindent Joining the expressions results in
        \begin{equation}\label{eq:xn}
            \matrixel{\psi}{\vu\rho_{X_n}}{\phi} = \sum_{i=1}^n p_{X_n}(x_i) \qty[e^{-|x_i|^2}\sum_{j,k=0}^\infty \psi_j^*\phi_k\frac{x_i^j(x_i^*)^k}{\sqrt{n!m!}}].
        \end{equation}
        
        Now, take the Glauber-Surdashan P-representation of a thermal state $\vu\rho^{th}(V_m)$ and compute $\matrixel{\psi}{\vu\rho^{th}(V_m)}{\phi}$:
        \begin{align}
            \matrixel{\psi}{\vu\rho^{th}(V_m)}{\phi} &= \sum_{j=0}^\infty\psi_j^*\bra{j}\qty[\int_\mathbb{C}p(x)\op{x}\dd[2]{x}]\sum_{k=0}^\infty\phi_k\ket{k}\\
            &= \int_\mathbb{C}p(x)\qty[\sum_{j,k=0}^\infty\psi_j^*\phi_k\braket{j}{x}\braket{x}{k}]\dd[2]{x}\\\label{eq:xg}
            &= \int_\mathbb{C}p(x)\qty[e^{-|x|^2}\sum_{j,k=0}^\infty\psi_j^*\phi_k\frac{x^j(x^*)^k}{\sqrt{j!k!}}]\dd[2]{x},
        \end{align}
        \noindent Now, as $X_n\rightarrow X_G$, one has that\footnote{Here, if $A$ is a random variable, $\mathcal{L}(A)$ represents the law of $A$.} $\mathcal{L}(X_n)\rightarrow\mathcal{L}(X_G)$ \cite{billingsley1999} and comparing \eqref{eq:xn} and \eqref{eq:xg}, $\matrixel{\psi}{\vu\rho_{X_n}}{\phi}\rightarrow\matrixel{\psi}{\vu\rho^{th}(V_m)}{\phi}$. Since $\ket{\psi}$ and $\ket{\phi}$ were arbitrarily chosen, we conclude that if $X_n\rightarrow X_G$ then $\vu\rho_{X_n}\rightarrow\vu\rho_{X_G}=\vu\rho^{th}(V_m)$.
    \end{proof}

    \begin{remark}
        Since the sequence $\qty{X_n}_{n\in\mathbb{N}}$ was chosen such that each element has the same first and second moments, one has $\vu\rho_{X_n}^G = \vu\rho_{X_G}^G = \vu\rho_{X_G} = \vu\rho^{th}(V_m)$.
    \end{remark}

    With the convergence of density operators towards a Gaussian quantum state defined, we can now explore the convergence of non-Gaussianity measure.
    
    \begin{corollary}\label{cor:ng-convergence}
        If $X_n\overset{\mathcal{D}}{\rightarrow}X_G$ then $\lim_{n\rightarrow\infty}\delta_{vN}(\vu\rho_{X_n}) = 0$.
    \end{corollary}
    
    \begin{proof}
        From \Cref{th:entropy-semicontinuity,th:state-convergence} we have $S(\vu\rho_{X_G})\leq \liminf_{n\rightarrow\infty} S(\vu\rho_{X_n})$ and from the Gaussian extremality theorem, $S(\vu\rho_{X_G})\geq S(\vu\rho_{X_n})$. Then, $\lim_{n\rightarrow\infty} S(\vu\rho_{X_n}) = S(\vu\rho_{X_G})$. Since $\delta_{vN}(\vu\rho_{X_n}) = S(\vu\rho_{X_n}^G) - S(\vu\rho_{X_n})$ and $\vu\rho_{X_n}^G = \vu\rho_{X_G}$, $\lim_{n\rightarrow\infty}\delta_{vN}(\vu\rho_{X_n}) = 0$.
    \end{proof}

    As defined, $\varepsilon_G(\vu\rho_{AB})$ represents the loss of SKR as a consequence of assuming a Gaussian model for the data observed by Alice and Bob. The amount of SKR lost is proportional to Bob's state nG, and, since the channel connecting Alice and Bob is unknown, it involves an optimization considering all quantum channels compatible with the observed statistics, usually the transmittance $\tau$ and the excess noise $\xi$. The following result states that this loss of SKR becomes negligible as the cardinality grows if the constellation shape is chosen properly.
    
    \begin{corollary}\label{cor:qkd-protocol-nG-convergence}
        If $X_n\overset{\mathcal{D}}{\rightarrow}X_G$ then $\varepsilon_G(\vu\rho_{AB})\rightarrow 0$. 
    \end{corollary}
    
    \begin{proof}
        From \Cref{lemma:qkd-protocol-ng}, the convergence of two quantities need to be analyzed. For the classical one, we have $D_{X_n}(snr)\rightarrow0$ since $X_n\overset{\mathcal{D}}{\rightarrow}X_G$ regardless of the quantum channel being considered because it must match the observed statistics during parameter estimation. We now must show that $\inf_{\mathcal{N}_{A'\rightarrow B}}\qty{\delta_{vN}(\mathcal{N}(\vu\rho_A))}\rightarrow0$ as well. Let $\vu\rho$ be an arbitrary quantum state, $\mathcal{N}^*$ be the quantum channel achieving the infimum and take an arbitrary Gaussian quantum channel $\mathcal{N}$, both being compatible with the estimated parameters. Then, using \Cref{prop:ng-monotonicity} of $\delta_{vN}$, 
        \begin{equation}
            \delta_{vN}(\vu\rho_A) \geq \delta_{vN}(\mathcal{N}(\vu\rho_A)) \geq \delta_{vN}(\mathcal{N}^*(\vu\rho_A)).
        \end{equation}
        \noindent By making $\vu\rho_A = \vu\rho_{X_N}$ and choosing a sequence such that $X_n\overset{\mathcal{D}}{\rightarrow}X_G$, \Cref{cor:ng-convergence} ensures that
        \begin{equation}
            \lim_{n\rightarrow\infty}\delta_{vN}(\mathcal{N}^*(\vu\rho_{X_N})) \leq \lim_{n\rightarrow\infty}\delta_{vN}(\vu\rho_{X_n}) = 0.
        \end{equation}
        \noindent Then, $\varepsilon_G(\vu\rho_{AB})\rightarrow 0$.        
    \end{proof}

    The operational interpretation is that if Alice chooses $X_i\in\qty{X_n}$ to define the amplitudes and probability distribution of its coherent states (the constellation) and such sequence of random variables converges towards a normal distribution, she and Bob can safely reconstruct the covariance matrix for the equivalent entangled-based protocol and be sure that the SKR lost due to the lower bound becomes negligible if the constellation cardinality is big enough.

    \subsection{Numerical Results}
    
    To explore the results presented in the last Section, we computed the value of $\delta_{vN}$ for two types of $QAM$-like constellations under a variety of situations. These QAM constellations are derived by the Cartesian product of real valued discrete random variables. The first family of random variables is the Gauss quadrature (GQ) type and is defined as follows. For a standard Gaussian density function $p_X(x) = \frac{1}{\sqrt{2\pi}}e^{-x^2/2}$, we denote the $m$-th Hermite polynomial by $H_m$,
    \begin{equation}
        H_m(x) = \frac{(-1)^m}{p_X(x)}\dv[m]{p_X(x)}{x}.
    \end{equation}
    The $m$ roots $\qty{x_{i,m}}$ of $H_m$ assume the role of a set of real amplitudes together with the weights 
    \begin{equation}\label{eq:const-gq}
        w_{i,m} = \frac{(m-1)!}{mH^2_{m-1}(x_{i,m})},
    \end{equation}
    \noindent which is a proper discrete probability distribution ($w_{i,m}>0$ and $\sum_iw_{i,m} = 1$) also known as Gauss quadrature. Then, the Gauss quadrature constellation (GQ) is composed of the amplitudes $\mathcal{A}_{GQ} = \qty{x_{i,m}}$ and the probability distribution $\mathcal{P}_{GQ} = \qty{w_{i,m}}$ of \eqref{eq:const-gq}.

    The second family of random variables is called the random walk (RW). For the normalized random walk        
    \begin{equation}
    X_m = \frac{1}{\sqrt{m-1}}\sum_{k=1}^{m-1}Z_k,
    \end{equation}        
    \noindent where {$Z_k$} are i.i.d. on $\qty{1,-1}$, one has the following convergence in distribution,
    \begin{equation}\label{eq:const-rw}
    X_m \overset{D}{=} \frac{2}{\sqrt{m-1}}\qty(B_m - \frac{m-1}{2}),
    \end{equation}        
    \noindent being $B_m \sim \operatorname{Bin}(m-1,1/2)$. Then, the $m$-point Random Walk (RW) constellation is defined by the random variable $X_m$, that is, the amplitude set $\mathcal{A}_{RW} = \qty{(2i-m+1)/\sqrt{m-1}}_{i=0}^{m-1}$  with probability distribution $\mathcal{P}_{RW} = \operatorname{Bin}(m-1,1/2)$. In both constellations, GH and RW, we set the desired variance by proportionally adjusting the amplitudes.

\begin{figure*}[!t]
    \centering
    \subfloat[]{\includegraphics[width=2.5in,page=1]{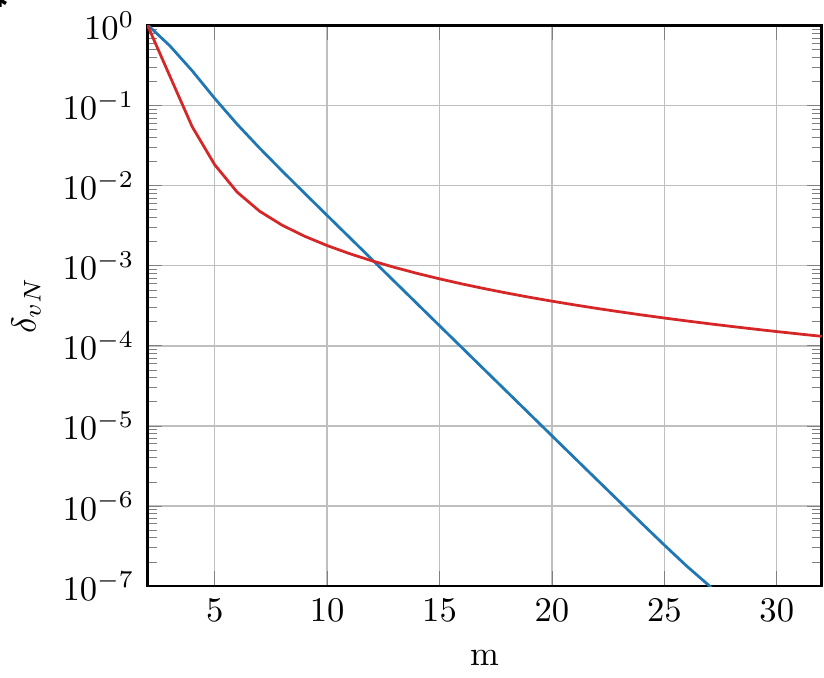}}%
    \subfloat[]{\includegraphics[width=2.5in,page=2]{plots/ng}}%
	\caption{(a) Values of the QRE-nG for the RW-QAM (blue line) and GH-QAM (red line) constellations with $N=m^2$ points and modulation variance $V_m=2.5$. (b) Values of the QRE-nG for the GH-QAM constellation with $N=m^2$ points under a thermal loss channel with fixed transmittance ($\tau = 0.5$). The upper line (blue) corresponds to the constellation nG prior to the channel and in the three lines below we set and increasing thermal noise $\bar{n} = \qty{0,0.2,0.4}$.}
	\label{fig:nG-const-thermal-chann}
	\end{figure*}

    In Fig.\ref{fig:nG-const-thermal-chann}a we plotted the values of $\delta_{vN}$ for constellations of coherent states with GH and RW shapes with fixed modulation variance $V_m=2.5$. The behavior of $\delta_{vN}$ seems to be very similar to the capacity gap of \Cref{def:capacity-gap} presented in \cite{wu2010} -- the RW constellation performs better than the GH for smaller $N$, and at $N=144$ it outperforms the RW. Also, it can be seen that, analogously to the classical capacity gap, the QRE-nG decreases exponentially, meaning that the non-Gaussian mixture of coherent states converges to the Gaussian equivalent state in an exponential pace.

\begin{figure*}[!t]
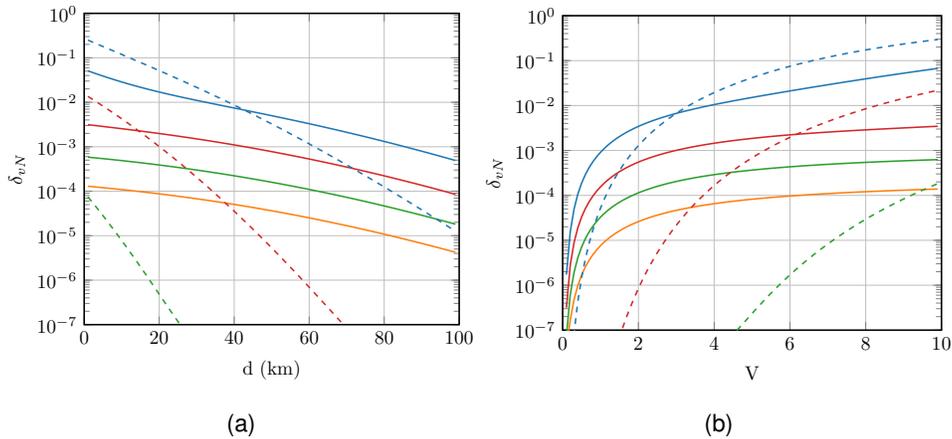

	\centering
	\subfloat[]{\includegraphics[width=2.5in,page=3]{plots/ng}}%
	\subfloat[]{\includegraphics[width=2.5in,page=4]{plots/ng}}%
	\caption{(a) Values of the QRE-nG for the RW-QAM (solid lines) and GH-QAM (dashed lines) constellations under a thermal loss channel with fixed thermal noise $\bar{n}=0.1$ and increassing transmittance $\tau = 10^{-0.01d}$, $d$ being the distance in kilometers. From top to bottom, the constellations sizes are 16, 64, 256 and 1024 points, respectively. (b) Values of the QRE-nG for the RW-QAM (solid lines) and GH-QAM (dashed lines) constellations as a function of the modulation variance under a thermal loss channel with fixed thermal noise $\bar{n}=0.1$ and transmittance $\tau = 10^{-0.01d}$ for $d = 50$km. From top to bottom, the constellations sizes are 16, 64, 256 and 1024 points, respectively.}
	\label{fig:ng-therm-loss-vs-dist_therm-loss-vs-variance}
\end{figure*}

    The plot in Fig.\ref{fig:nG-const-thermal-chann}b shows how $\delta_{vN}$ decreases under the action of the thermal loss quantum channel when the input is a $m^2$-QAM constellation with $2\geq m \geq 32$ (from a 4-QAM to a 1024-QAM). The upper blue line corresponds to the constellation nG at the channel input, and the lower lines are the output state nG for fixed transmittance $\tau=0.5$ and $\bar{n} = \{0,0.2,0.4\}$, respectively. It seems that the output state nG is monotone with respect to both the channel transmittance and the excess noise. This apparent monotone behavior can be seen in Fig.\ref{fig:ng-therm-loss-vs-dist_therm-loss-vs-variance}a where we plotted $\delta_{vN}$ as a function of distance and in Fig.\ref{fig:ng-therm-loss-vs-dist_therm-loss-vs-variance}b as a function of the modulation variance for both constellation shapes with sizes 16, 64, 256 and 1024. In both plots we see the consonance with the results observed in Fig.\ref{fig:nG-const-thermal-chann}a where the best constellation shape depends on the situation: in general, GH constellations perform better in longer distances as well as for bigger constellations; the 256GH-QAM constellation presents less non-Gaussianity than the 256RW-QAM for every distance and modulation variance considered.

    It is relevant to note that the measure of nG quickly decreases as the constellation size grows, even when nG is computed at the output of the quantum channel. Also, the von Neumann nG measure increases with the constellation average energy, which is expected as it becomes more easy to distinguish between the states in a set of distant coherent states when those states are more distant (in a phase space picture).
    
%

%% file: files/SEC06.tex
\section{Conclusion}\label{sec:conclusion}

In this paper, we explored the convergence of constellations of coherent states towards a reference Gaussian quantum state and its applications to the security analysis of discrete modulated CVQKD protocols. We defined a measure of non-Gaussianity of a DM-CVQKD protocol as the difference of secret key rate from the actual protocol to the Gaussian equivalent one providing a lower bound on the SKR, which is interpreted as the amount of SKR lost due to using a Gaussian model to provide a lower bound to protocol with non-Gaussian modulation.

We proved that the mixed state representing a constellation of coherent states converges weakly to a thermal state as the constellation size increases if the random variables describing the constellations form a sequence converging in distribution to a normal random variable. This fact was used to demonstrate that the constellation measure of non-Gaussianity vanishes with the increase of the constellation size, implying that if the DM-CVQKD protocol uses a constellation big enough, the loss of SKR due to the lower bound is negligible: Using a 256-QAM with Gauss-Hermite shaping, the loss of SKR quickly falls below $10^-5$ as the distance increases.

One of the main concepts used was the measure of non-Gaussianity of a quantum state based on the quantum relative entropy of von Neumann, which has the property of being monotone decreasing under Gaussian evolutions. We showed the existence of non-Gaussian quantum channels for which the QRE-nG is nonincreasing and provided the conditions under which it can be affirmed. 

Future work can focus on two main research topics. The first is an investigation of the implications of weak convergence for unconditional security of the DM-CVQKD protocols. Unconditional security refers to the relationship between security under collective attacks and security under arbitrary attacks. For CVQKD with Gaussian modulation, several frameworks have been proposed using de Finetti-type theorems or postselection. The second topic concerns the study of non-Gaussianity for quantum channels. In this paper, we used the notion of an nG measure of a quantum state and used it to analyze the DM-CVQKD. As recent works have been engaged to the development of non-Gaussian resource theory for quantum operations, it could be interesting how those tools can be applied to the non-Gaussian quantum channel relevant to QKD protocols.

%% file: files/SEC05.tex
\section{QRE non-Gaussianity Monotone under nG Quantum Channels}\label{sec:nG-monotone}

In the Proof of \Cref{cor:qkd-protocol-nG-convergence}, it was observed that, given a quantum state $\vu\rho$, it is possible that there exist non-Gaussian quantum channels such that $\delta_{vN}(\vu\rho) \geq \delta_{vN}(\mathcal{N}(\vu\rho))$, or at least the ones for which the QRE-nG measure does not increase. In other words, the monotone property of the QRE-nG may be extended to some set of quantum channels further than the Gaussian sector, even if it is restricted to some specific set of quantum states. The aim of this section is to describe the conditions for which a nG quantum channel maintains the monotonic property of QRE-nG and how such channels can be useful in the description of the relevant quantum channels in a DM-CVQKD protocol.
We begin by proving the following lemma.

\begin{lemma}\label{lemma:delta-channel-state}
    Let $\mathcal{N}$ be a quantum channel, $\vu\rho$ an arbitrary quantum state and define 
    \begin{equation}
        \Delta(\mathcal{N}, \vu\rho) = \tr[\mathcal{N}(\vu\rho)(\log\mathcal{N}(\vu\rho)^G - \log\mathcal{N}(\vu\rho^G))].
    \end{equation}
    \noindent If $\mathcal{N}$ is Gaussian, then $\Delta(\mathcal{N}, \vu\rho) = 0$ for any quantum state $\vu\rho$.
\end{lemma}
\begin{proof}
    If $\mathcal{N}$ is a Gaussian channel and $\bm\Gamma$ is the covariance matrix of an arbitrary quantum state $\vu\rho$, then $\bm\Gamma(\vu\rho) = \bm\Gamma(\vu\rho^G)$ and $\bm\Gamma\overset{\mathcal{N}}{\rightarrow}\bm\Gamma'$. This means that $\bm\Gamma(\mathcal{N}(\vu\rho)^G) = \bm\Gamma(\mathcal{N}(\vu\rho^G)) = \bm\Gamma'$. Since the first moment will follow as same, $\mathcal{N}(\vu\rho^G) = \mathcal{N}(\vu\rho)^G$ for any $\vu\rho\in\mathcal{D}(\mathcal{H})$ and then $\Delta(\mathcal{N}, \vu\rho)=0$ for arbitrary $\vu\rho$.
\end{proof}


This result gives a sufficient condition to classify a quantum channel with respect to its non-Gaussianity: if the functional $\Delta(\mathcal{N},\vu\rho)\neq0$ for any quantum state $\vu\rho$, $\mathcal{N}$ is nG. If we define $\mathcal{F}=\qty{\mathcal{N}\in\mathcal{Q}: \Delta(\mathcal{N}, \vu\rho)\geq 0 \,\forall\,\vu\rho\in\mathcal{D}(\mathcal{H})}$, we have $\mathcal{G}\subset\mathcal{F}$ and we are able to prove the following statement.

\begin{theorem}\label{th:nG-chan-mon-dec}
    If $\mathcal{N}\in\mathcal{F}$ then $\delta_{vN}(\mathcal{N}(\vu\rho))\leq\delta_{vN}(\vu\rho)$ for any $\vu\rho\in\mathcal{D}(\mathcal{H})$.
\end{theorem}

\begin{proof}
    Let $\vu\rho$ and $\mathcal{N}$ be as in the setup. From quantum relative entropy contractivity under quantum channels, one gets
    \begin{align}
        \delta_{vN}(\vu\rho) &\overset{(a)}{=} S(\vu\rho||\vu\rho^G)\\
             &\overset{(b)}{\geq} S(\mathcal{N}(\vu\rho)||\mathcal{N}(\vu\rho^G))\\
            &\overset{(c)}{=}  \tr[\mathcal{N}(\vu\rho)(\log\mathcal{N}(\vu\rho) - \log\mathcal{N}(\vu\rho^G))] + \tr[(\mathcal{N}(\vu\rho) - \mathcal{N}(\vu\rho)^G)\log\mathcal{N}(\vu\rho)^G ]\\
            &\overset{(d)}{=} S(\mathcal{N}(\vu\rho)^G) - S(\mathcal{N}(\vu\rho)) + \Delta(\mathcal{N}, \vu\rho)\\
            &\overset{(e)}{=} S(\mathcal{N}(\vu\rho)||\mathcal{N}(\vu\rho)^G) + \Delta(\mathcal{N}, \vu\rho)\\
             &\overset{(f)}{\geq} \delta_{vN}(\mathcal{N}(\vu\rho)),
    \end{align}
    \noindent where (a) comes from the definition of $\delta_{vN}$, (b) from the monotonicity of quantum relative entropy \cite{wilde2017a}, (c) one has that $\tr[(\vu\sigma - \vu\sigma^G)\log(\vu\sigma^G)] = 0$ for arbitrary $\vu\sigma$ \cite{holevo1999}, (d) we used the definition in \Cref{lemma:delta-channel-state}, (e) from \Cref{def:qre-ng} and (e) because $\mathcal{N}$ and $\vu\rho$ where chosen such chat $\Delta(\mathcal{N}, \vu\rho) \geq 0$.
\end{proof}

The above result extends the monotone property of $\delta_{vN}$ under Gaussian quantum channels to nG channels and also provides an interpretation of the quantity given by $\Delta(\mathcal{N}, \vu\rho)$: if it is nonnegative for every $\vu\rho$, $\mathcal{N}$ does not increase QRT-nG. It should be noted that the difference $S(\vu\rho||\vu\rho^G) - S(\mathcal{N}(\vu\rho)||\mathcal{N}(\vu\rho^G))$ is related to state recovery maps (Petz recovery maps), which are maps that can recover the state that suffered some physical evolution. We also stress that recovery maps can be extended to quantum systems in infinite dimensions \cite{junge2018}.

However, the specification of $\mathcal{F}$ may have been too broad by demanding $\Delta(\mathcal{N}, \vu\rho)$ to be nonnegative for all quantum states in the system and we cannot affirm whether $\mathcal{F}\setminus\mathcal{G} = \qty{\varnothing}$ or not. A relaxation in this condition can be done by considering only a specific set of quantum states, which we chose to be the states relevant to the context of DM-CVQKD protocols, and can be helpful in describing a set of QRE-nG nonincreasing quantum channels. 

Let $\mathcal{S}_{\bar{n}} = \{\vu\sigma\in\mathcal{D}(\mathcal{H}):\vu\sigma = \sum_{x\in\mathcal{X}_n}p(x)\vu\rho^{th}(x, \bar{n})\}$ with $X_n$ being a discrete symmetric random variable and $\vu\rho^{th}(x, \bar{n})$ be the displaced thermal state with $\bar{n}$ average photons and the first moment $\bar{\bm{x}} = 2\cdot(\Re{x}, \Im{x})^T$. Constellations of coherent states are represented by the set $\mathcal{S}_0$ and any state in $\mathcal{S}_{\bar{n}}$ has a diagonal covariance matrix for any value of $\bar{n}$, which means that its equivalent Gaussian quantum state is a thermal state with the appropriate mean photon number. Then, we can define a relaxed set $\mathcal{F}_{\bar{n}}=\qty{\mathcal{N}\in\mathcal{Q}: \Delta(\mathcal{N}, \vu\rho)\geq 0 \,\forall\,\vu\rho\in\mathcal{S}_{\bar{n}}}$ such that the QRE-nG of any quantum state in $\mathcal{S}_{\bar{n}}$ is nonincreasing under the action of any channel in $\mathcal{F}_{\bar{n}}$. Also, we have $\mathcal{G}\subset\mathcal{F}\subset\mathcal{F}_{\bar{n}}$ for any $\bar{n}$. The states in $\mathcal{S}_{\bar{n}}$ are relevant for the QKD setup because they represent the mixed states output by a thermal loss channel with thermal noise $\bar{n}$. We can affirm the following proposition.

\begin{proposition}\label{prop:ng-chan}
    $\mathcal{F}_0\setminus\mathcal{G} \neq \qty{\varnothing}$.
\end{proposition}
\begin{proof}
    Take the phase diffusion process described in Appendix \ref{ap:phase-diff-chan} and represented by $\mathcal{N}_\Delta$, which is the model of a non-Gaussian evolution of a quantum system. It is known that the QRE-nG of coherent states under phase diffusion increases with time and in Appendix \ref{ap:phase-diff-chan} we show that for any $\vu\rho\in\mathcal{S}_0$, $\bar{\bm{x}}(\vu\rho) =  \bar{\bm{x}}(\mathcal{N}_\Delta(\vu\rho))$ and $\bm\Gamma(\vu\rho) = \bm\Gamma(\mathcal{N}_\Delta(\vu\rho))$, which implies that $\vu\rho^G = \mathcal{N}_\Delta(\vu\rho)^G$. That is, the phase diffusion process does not modify the first and second statistical moments of appropriated mixtures of coherent states. In addition, it does not have an effect on thermal states, meaning that $\mathcal{N}_\Delta(\vu\rho^G) = \vu\rho^G$. We conclude that $\mathcal{N}_\Delta(\vu\rho^G) = \mathcal{N}_\Delta(\vu\rho)^G$ which results in $\Delta(\mathcal{N}_\Delta,\vu\rho)=0$ for any state in $\mathcal{S}_0$ and then in $\mathcal{F}_0\setminus\mathcal{G} \neq \qty{\varnothing}$.
\end{proof}

\begin{figure}[!tb]
    \centering
    \includegraphics[width=2.5in,page=5]{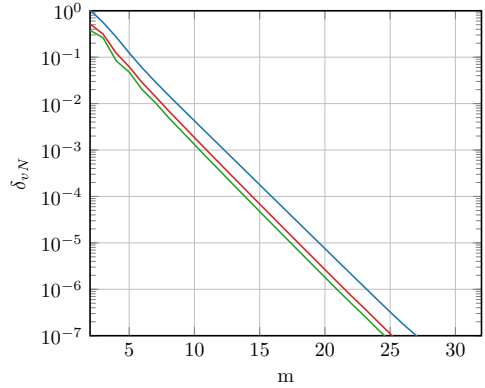}
    \caption{\label{fig:ng-phase-diff-chan}Values of the QRE-nG for the GH-QAM constellation under a phase diffusion process with fixed modulation variance $V_m=2.5$ and increasing constellation size. The upper line (blue) corresponds to the constellation nG prior to the channel and in the constelaltion under the process with parameters $\gamma=0.15$ and $\gamma=\infty$, respectively.}
\end{figure}

We conjecture that \Cref{prop:ng-chan} can be extended to other values of $\bar{n}$ different from zero, although we have not yet worked out the proof. We plotted in Fig.\ref{fig:ng-phase-diff-chan} the QRE-nG of constellations of coherent states that undergo the phase diffusion process. The upper blue line corresponds to the QRE-nG measure for the constellations prior to the process taking place, and we calculated $\delta_{vN}(\mathcal{N}_\Delta(\vu\rho_{X_n}))$ for $\lambda=0.15$ (red line) and $\lambda=\infty$ (green line), which have the effect of total decoherence in the mixture of coherent states, destroying the off-diagonal elements in the density matrix. As expected, the action of $\mathcal{N}_\Delta$ does not increase the QRE-nG of the constellation once $\vu\rho_{X_n}\in\mathcal{S}_0$.

In addition to the fact that the phase diffusion process preserves the monotone property of the QRE-nG when restricted to an appropriate set of quantum states, the fact that it leaves the covariance matrix of the constellation invariant is of an interesting implication to the DM-CVQKD protocols. Define by $\mathcal{T}$ the set of quantum channels that preserve the first and second moments of any quantum state in $\mathcal{S}_{\bar{n}}$ for any value of $\bar{n}$. Now, take a quantum thermal loss channel $\mathcal{N}_1$ with transmittance $\tau$ and excess noise $\xi$ and any quantum channel $\mathcal{N}_2\in\mathcal{T}$. In the parameter estimation step of a DM-CVQKD protocol, Alice and Bob use part of the data to estimate the channel transmittance and excess noise in order to reconstruct the covariance matrix of the bipartite state in the entangled-based protocol. This covariance matrix is used to compute the bounds on the Holevo information for the eaversdropper information, and every quantum channel compatible with the estimated parameters must be considered.

The idea here is that the quantum channels considered in a DM-CVQKD security analysis can be broken down into two channels, the Gaussian $\mathcal{N}_1$ yielding the observed parameters and $\mathcal{N}_2$, which does not modify the covariance matrix (and then does not affect the parameter estimation), but is responsible to non-Gaussian interactions that are referenced to the information obtained by the eavesdropper.

%% file: files/appendix.tex
\section{The Phase-Diffusion Process and its Effect on Gaussian States}\label{ap:phase-diff-chan}

One of the most harmful noisy processes in quantum information is the one that provokes decoherence, where the quantum states lose their \aspas{quantumness} ---- the full decohered states reduce to mixtures of orthogonal states which can be perfectly distinguished. Some open system processes result in random changes in the relative phase of the states that are in superposition in the main quantum system. Such a relative phase fluctuation results in a loss of coherence and is called phase damping or phase diffusion \cite{liu2004}. 

The single mode evolution of the system under such process can be described by the master equation \cite{genoni2010,memarzadeh2016}
\begin{equation}
    \dv{t}\vu\rho = \Gamma\mathcal{L}[\opad\opa]\vu\rho,
\end{equation}
\noindent where $\mathcal{L}[\vu{O}]\vu\rho = 2\vu{O}^\dagger\vu\rho\vu{O} - \vu{O}^\dagger\vu{O}\vu\rho - \vu\rho\vu{O}^\dagger\vu{O}$, or as the Hamiltonian of a harmonic oscillator open to an $N$ mode environment\cite{liu2004}
\begin{equation}
    H = \hbar\omega\opad\opa + \hbar\sum_{i=1}^N\omega_i\opad_i\opa_i+\hbar\sum_{i=1}^N\chi_i\opad\opa(\opa_i+\opad_i),
\end{equation}
\noindent where $\opa$ and $\opad$ are the anihilation and creation operators of the main system with frequency $\omega$ and $\opa_i$ and $\opad_i$ refer to the $i$th environment system with frequency $\omega_i$. The quantity $\chi_i$ represents a coupling parameter between the main system and the $i$th environment mode.

This non-Gaussian evolution of a quantum state is an important source of noise in optical communication links. Its Krauss operator set $\{P_k(t)\}$, $0\leq k\leq\infty$ has elements
\begin{equation}
    P_k(t) = \sum_{n=0}^\infty e^{-\frac12n^2\lambda^2}\sqrt{\frac{(n^2\lambda^2)^k}{k!}}\op{n}
\end{equation}
\noindent where $\lambda = t\sqrt{\Lambda}$ and $\Lambda = \sum_i\chi_i^2\sqrt{1-e^{-n^2\lambda^2}}$.

Let us first inspect the phase diffusion evolution of a thermal state with average photon number $\bar{n}$, $\vu\rho^{th}(\bar{n})$:
\begin{align}\nonumber
    \sum_{k=0}^\infty P_k(t)\vu\rho^{th}(\bar{n})P_k^\dagger(t) =& \sum_{k,n=0}^\infty e^{-\frac12n^2\lambda^2}\sqrt{\frac{(n^2\lambda^2)^k}{k!}}\op{n} \cdot \sum_{m=0}^\infty\frac{\bar{n}^m}{(\bar{n}+1)^{m+1}}\op{m}\times\\
     &\,\sum_{l=0}^\infty e^{-\frac12l^2\lambda^2}\sqrt{\frac{(l^2\lambda^2)^k}{k!}}\op{l},\\
    =& \sum_{k,l,m,n=0}^\infty \frac{e^{-\frac12n^2\lambda^2}e^{-\frac12l^2\lambda^2}\bar{n}^m\sqrt{(n^2\lambda^2)^k(l^2\lambda^2)^k}}{k!(\bar{n}+1)^{m+1}}\ket{n}\braket{n}{m}\braket{m}{l}\bra{l}\\
    =& \sum_{k=0}^\infty\sum_{n=0}^\infty e^{-n^2\lambda^2}\frac{(n^2\lambda^2)^k}{k!}\frac{\bar{n}^n}{(\bar{n}+1)^{n+1}}\op{n}\\
    =& \sum_{n=0}^\infty e^{-n^2\lambda^2}\sum_{k=0}^\infty\qty[\frac{(n^2\lambda^2)^k}{k!}]\frac{\bar{n}^n}{(\bar{n}+1)^{n+1}}\op{n}\\
    =& \sum_{n=0}^\infty\frac{\bar{n}^n}{(\bar{n}+1)^{n+1}}\op{n} = \vu\rho^{th}(\bar{n})
\end{align}

Then, the thermal states are invariant under the phase diffusion process. This behavior is not present in all Gaussian states, once for a coherent state $\ket{\alpha}$ with $\alpha = (q+ip)/2$,
\begin{equation}\label{eq:coh-phase-dif-state}
    \begin{aligned}[b]    
        \mathcal{N}_\Delta(\op{\alpha}) =& e^{-|\alpha|^2}\sum_{m,n=0}^\infty \exp{-\Delta^2(n-m)^2}\\
         &\times\frac{\alpha^n\alpha^{*m}}{\sqrt{n!m!}}\op{n}{m}
    \end{aligned}    
\end{equation}
\noindent is a non-Gaussian mixed state with $\Delta = \lambda^2/2$. To derive the expressions of the mean vector and the covariance matrix of $\mathcal{N}_\Delta(\op{\alpha})$, we need the expectations of the operators $\opa$, $\opad$, $\opa^2$, $\opa^{\dagger2}$ and $\opad\opa$ which can be computed using the expression in \eqref{eq:coh-phase-dif-state} and the formulation in Sec. \ref{sec:gaussian-system}, resulting in
\begin{align*}
    \ev{\opa} &= e^{-\Delta^2}\alpha,  &\ev*{\opad}&= e^{-\Delta^2}\alpha^*,  &\ev*{\opa^2} &= e^{-4\Delta^2}\alpha^2\\
    \ev*{\opa^{\dagger2}} &= e^{-4\Delta^2}\alpha^{*2},  &\ev*{\opad\opa} &= e^{-\Delta^2}|\alpha|^2. &
\end{align*}

So, the mean vector is
\begin{equation}
    \bar{\bm{x}} = \mqty(\ev{\opq} \\ \ev{\opp}) = \mqty( q\cdot e^{-\Delta^2} \\ p\cdot e^{-\Delta^2} ),
\end{equation}
\noindent and the covariance matrix elements are
\begin{align}
    [\bm\Gamma]_{1,1} &= 1 + 2|\alpha|e^{-\Delta^2} + e^{-4\Delta^2}(\alpha^2 + \alpha^{*2}) - q^2e^{-2\Delta^2}\\
    [\bm\Gamma]_{2,2} &= 1 + 2|\alpha|e^{-\Delta^2} - e^{-4\Delta^2}(\alpha^2 + \alpha^{*2}) - p^2e^{-2\Delta^2}\\
    [\bm\Gamma]_{1,2} &= [\bm\Gamma]_{2,1} = qp(e^{-4\Delta^2} - e^{-2\Delta^2})
\end{align}

In contrast to thermal states, coherent states suffer from decoherence in a phase-diffusion evolution, with the off-diagonal elements of the density operator being more affected as the noise parameter $\Delta$ becomes larger. In addition, the mean vector and the covariance matrix also change with $\Delta$. 

Now, let us consider mixtures of coherent states, specifically those representing proper digital modulation constellations, such as $\vu\rho = \sum_ip_i\op{\alpha_i}$, and denote $\vu\rho_\Delta = \mathcal{N}_\Delta(\vu\rho)$. Due to the linearity of the channel, we use \eqref{eq:coh-phase-dif-state} directly and have
\begin{equation}\label{eq:mix-phase-dif-state}
        \vu\rho_\Delta = \sum_{m,n=0}^\infty e^{-\Delta^2(n-m)^2}\sum_{i=0}^Np_ie^{-|\alpha_i|^2}\frac{\alpha_i^n\alpha_i^{*m}}{\sqrt{n!m!}}\op{n}{m}
\end{equation}

Clearly, as each coherent state is transformed into an nG mixed state, the resultant mixture will also be nG. Deriving the expectations of bosonic operators gives the following.
\begin{align}
    \tr(\opa\vu\rho_\Delta) &= \sum_{i=1}^N\Theta_i\alpha_i & \tr(\opad\vu\rho_\Delta) = \sum_{i=1}^N\Theta_i\alpha_i^*,
\end{align}
\noindent where we made $\Theta_i = e^{-\Delta^2}p_ie^{-|\alpha_i|^2}\cosh{|\alpha_i|^2}$ resulting in that
\begin{equation}
    \tr(\opq\vu\rho_\Delta) = \tr(\opp\vu\rho_\Delta) = 0,
\end{equation}
\noindent meaning that the phase diffusion process does not modify the first moment. For the second moment, we have the following quantities,
\begin{align}
    \tr(\opa^2\vu\rho_\Delta) &= e^{-4\Delta^2}\sum_{i=1}^Np_i\alpha_i^2\\
    \tr(\opa^{\dagger2}\vu\rho_\Delta) &= e^{-4\Delta^2}\sum_{i=1}^Np_i\alpha_i^{*2}\\
    \tr(\opad\opa\vu\rho_\Delta) &= \tr(\opad\opa\vu\rho) = \sum_{i=1}^Np_i|\alpha_i|^2,
\end{align}
\noindent giving the elements of the covariance matrix,
\begin{align}
    [\bm\Gamma]_{1,1} &= [\bm\Gamma]_{2,2} = 1 + 2\sum_{i=1}^Np_i|\alpha_i|^2\\
    [\bm\Gamma]_{1,2} &= [\bm\Gamma]_{2,1} = 0
\end{align}
\noindent so that $\bm\Gamma(\vu\rho_\Delta) = \bm\Gamma(\vu\rho)$. In what was exposed, the phase diffusion channel, not being Gaussian, does not necessarily map Gaussian states into Gaussian states, but there are cases in which this happens: thermal states are invariant under such physical process, in opposition to coherent states. Interestingly, if one prepares \aspas{symmetric} mixtures (with respect to the origin in the phase space) of coherent states, which already is not Gaussian so that the process would not \aspas{map it back} do Gaussianity, its first and second moments are preserved under phase diffusion evolution. We can then assume that there is a set of quantum states (convex mixtures of coherent states) whose first and second statistical moments are invariant under some non-Gaussian evolutions of the system. 

\section{Computing von Neumann Entropy from the Gram Matrix}\label{ap:ent-gram-matrix}

The von Neumann entropy of a quantum state $\vu\rho$ is defined as $S(\vu\rho) = -\tr(\vu\rho\log\vu\rho)$ and, if $\vu\rho$ has an eigendecomposition $\vu\rho = \sum\lambda_i\op{\lambda_i}$, $S(\vu\rho) = H(\qty{\lambda_i}) = -\sum\lambda_i\log(\lambda_i)$, where $H(\qty{\lambda_i})$ is the Shannon entropy of the set of eigenvalues of $\vu\rho$ ($\sum\lambda_i=1$). Computing $S(\vu\rho)$ can become challenging depending on the state $\vu\rho$.

\begin{definition}[Normalized Gram Matrix]\label{def:norm-gram-matrix}
	Let $\mathcal{E} = \qty{\ket{\psi_i}, p_i}_{i=1}^n$ be an ensemble of pure states. We call the normalized Gram Matrix $bm{G}$ the matrix with elements \begin{equation}
		\qty[\bm{G}]_{m,n} = \sqrt{p_mp_n}\braket{\psi_m}{\psi_n}.
	\end{equation}
\end{definition}

\begin{theorem}[Entropy of a Normalized Gram Matrix]\label{th:entropy-gramm}
	For an ensemble of pure states $\mathcal{E} = \qty{\ket{\psi_i}, p_i}_{i=1}^n$ in a quantum system of $d$ dimensions, being $\vu\sigma = \sum_ip_i\op{\psi_i}$ the density operator corresponding to the mixture, on has that the Gramm matrix $\bm{G}$ of $\mathcal{E}$ has the following properties:
	\begin{enumerate}
		\item $\bm{G}$ and $\vu\sigma$ have the same non-negative eigenvalues, including the same multiplicity. For a general case, where $n\neq d$, the issue of the quantity of eigenvalues is solved by including null eigenvalues. The consequence is that $\bm{G}$ and $\vu\sigma$ have the same von Neumann entropy, that is, $S(\vu\sigma) = S(\bm{G})$.
		
		\item $\bm{G}$ is positive and has unit trace. Moreover, if an arbitrary $m\times m$ matrix $\bm{A}$ is positive and has unit trace, then it is the Gramm matrix of a mixture of $m$ states in a $m$ dimensional quantum system.
	\end{enumerate}
\end{theorem}

\Cref{th:entropy-gramm} presents a quick method for computing the entropy of arbitrary mixtures of pure states. More importantly, the entropy of ensembles of CV pure quantum states is straightforward. For example, the entropy of Alice's modulation scheme can be computed by the corresponding Gram matrix as well as the entropy of the output of a loss-only quantum channel. Unfortunately, one cannot use \Cref{th:entropy-gramm} for arbitrary mixed states. In fact, the Hilbert-Schimidt inner product of the density matrices does not preserve the phase information. 